\documentclass[12pt]{article}
\usepackage{jeosman}
\usepackage{epsfig}
\usepackage{amsmath}
\usepackage{amssymb}

\begin{document}

\title{Enhanced transmission through arrays of subwavelength holes in gold films coated by a finite dielectric layer}

\author{Sanshui Xiao and Niels Asger Mortensen}
\address{MIC -- Department of Micro and Nanotechnology, NanoDTU,\\Technical
University of Denmark, DK-2800 Kongens Lyngby, Denmark}
\email{sanshui.xiao@mic.dtu.dk,nam@mic.dtu.dk}

\author{Min Qiu}
\address{Laboratory of Optics, Photonics and Quantum
Electronics, Department of Microelectronics and Applied Physics,
Royal Institute of Technology (KTH), Electrum 229, 16440 Kista,
Sweden.}


\begin{abstract}
Enhanced transmissions through a gold film with arrays of
subwavelength holes are theoretically studied, employing the rigid
full vectorial three dimensional finite difference time domain
method. Influence of air-holes shape to the transmission is
firstly studied, which confirms two different resonances
attributing to the enhanced transmission: the localized waveguide
resonance and periodic surface plasmon resonances. For the film
coated with dielectric layers, calculated results show that in the
wavelength region of interest the localized waveguide resonant
mode attributes to sensing rather than the periodic gold-glass
surface plasmon mode. Although the detected peak is fairly broad
and the shift is not too pronounced, we emphasize the contribution
for sensing from the localized waveguide resonant mode, which may
opens up new ways to design surface plasmon based sensors.
\end{abstract}

\maketitle
\section{Introduction}
In the recent years, the demonstration of a strong and unexpected
enhancement of light transmission through arrays of subwavelength
holes has generated numerous experimental and theoretical
work~\cite{Ebbesen:1998,Ghaemi:1998,Porto:1999,Martin:2001,Krishnan:2001,Klein:2004,Van:2005,Garcia:2005,Ruan:2006}.
Although there are still continuing discussions about the physical
mechanisms responsible for the extraordinary transmission, many
researchers show that the enhanced transmission is attributed to
the resonances. A surface plasmon (SP) is a collective oscillation
of free electrons inside a metal-dielectric surface. Strong
confinement of plasmonic waves near the metal surface provides
possibility for enhancing interactions with an analyte layer and
thus for efficient sensing of surface binding events. Surface
plasmon (SP) sensors are widely used in chemical and biological
research~\cite{Homola:1999}. Conventional SP sensors, employing
Kretschmann configuration, usually operate in SP-based attenuation
total internal refraction. Recently, there has been a growing
interest in surface plasmon resonance (SPR) sensing of
biochemicals using nanohole
arrays~\cite{Brolo:2004,Rindzevicius:2005,Tetz:2006}. These SP
sensors are based on the enhanced transmission through arrays of
nanoholes. It is a general consensus that these sensors are based
on SPRs, while the particular nature of the SPR and the mechanism
behind the sensing is still a questing open for discussion. In
this paper, we will theoretically study the transmission through
arrays of nanoholes and point out the contribution for sensing
from the localized waveguide resonant mode utilizing the rigid
full-vectorial three-dimensional (3D) finite-difference
time-domain (FDTD) method.

\section{Calculations and discussion}
\begin{figure}[t!]
\begin{center}
\epsfig{file=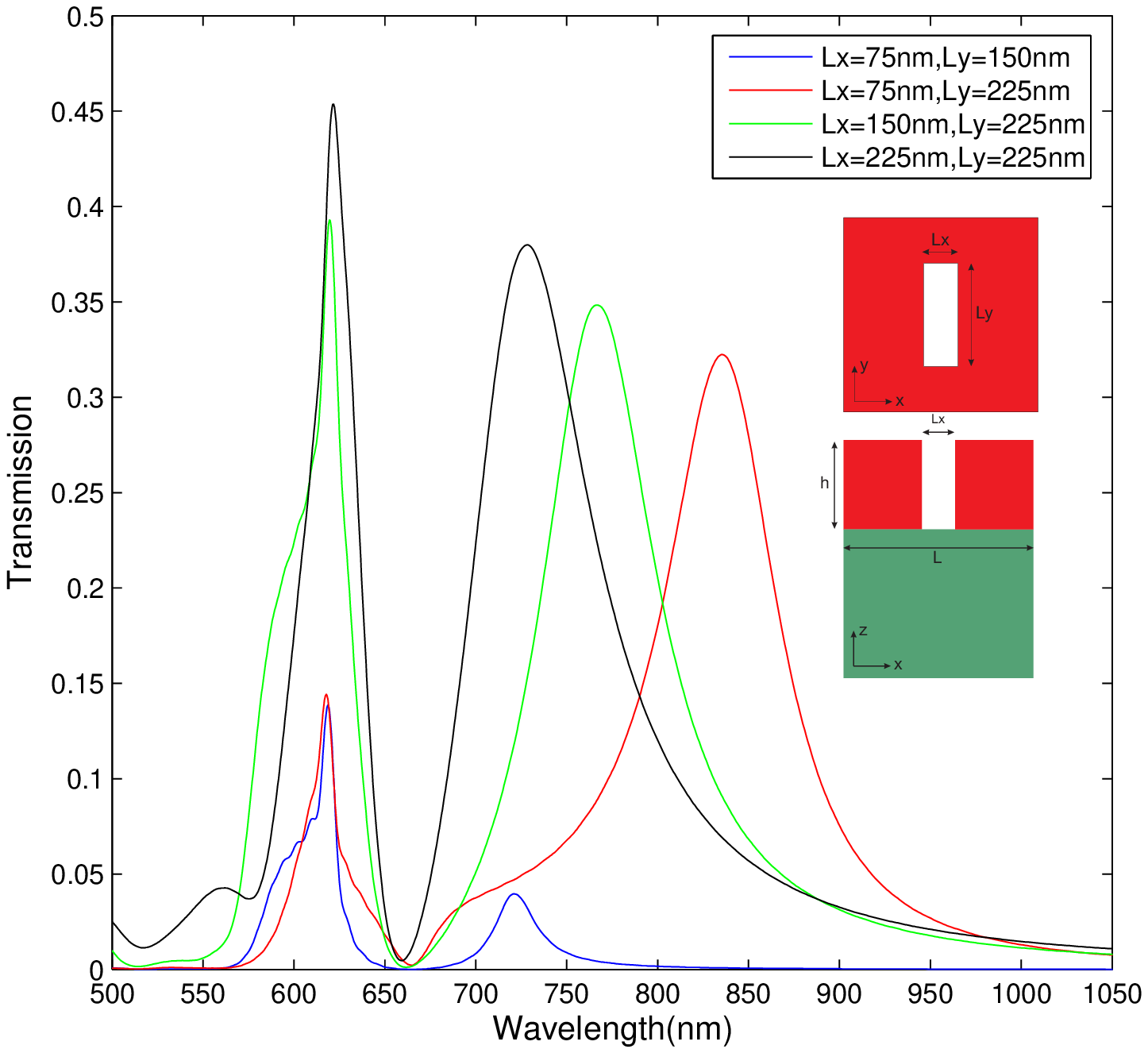,width=1 \columnwidth,clip,angle=0}
\end{center}
\caption{Transmission through the Au films with periodic square
arrays of aperture with different air-holes shape. The Au film is
on a glass substrate. The dimension of the holes is denoted by
Lx$\times$ Ly, , the thickness of the film is 200 nm and the
lattice constant (L) is 425 nm. } \label{shapes}
\end{figure}

Consider a gold (Au) film patterned with a periodic square array
of rectangular air holes. The Au film, assumed to be on a glass
substrate ($\varepsilon$=2.117), has a thickness of h=200 nm in
the z direction, the dimensions of the holes in xy directions are
denoted by Lx $\times$ Ly, and the lattice constant is denoted by
L. The insets in Fig. \ref{shapes} show the corresponding
structure with cross sections in the xy and yz planes, where the
green region represents the substrate while red for the metal. For
a similar structure, it has been pointed
out~\cite{Garcia:2005,Ruan:2006} that there are two different
resonances attributing to the enhanced transmission: (i) localized
waveguide resonances where each air hole can be considered as a
low-quality-factor resonator, and (ii) well-recognized surface
plasmon resonances due to the periodicity. These results have been
explained well by the band structure theory~\cite{Ruan:2006}. To
further support these explanation, here we study the transmission
through the Au film with periodic square arrays of different
air-holes shape. Consider the case of normal incidence, and the
electric field of the incident wave is polarized along the short
edge of the rectangular holes (the x direction). Transmission
through the film is calculated by the time-domain auxiliary
differential equation approach combining 3D FDTD models of a
dispersive material~\cite{TafloveFDTD,Xiao:2006c}. The dielectric
function of Au is described by the lossy Drude model
\begin{eqnarray}
\varepsilon(\omega)=\varepsilon_\infty-\frac{(\varepsilon_0-\varepsilon_\infty)\omega_p^2}{\omega^2+2i\omega\nu_c},
\end{eqnarray}
where $\varepsilon_\infty$ and $\varepsilon_0$ are the relative
permittivities at infinite and zero frequency, respectively,
$\omega_p$ is the plasma frequency, and $\nu_c$ is the collision
frequency. We choose $\varepsilon_\infty$=12.18,
$\varepsilon_0$=12.75, $\omega_p$=1.916$\times 10^{16}$rad/s, and
$\nu_c=1.406\times 10^{14}$rad/s for the Drude model, which fits
the experimental data~\cite{Palikbook} quite well.

Figure~\ref{shapes} shows the transmission through the gold film
with a square array of rectangular holes with different sizes in
the xy directions. The peak in the transmission at the wavelength
$\lambda=617 nm$ hardly moves with varying the air-holes shape,
while, the position of another peak is strongly dependant on the
air-holes shape. For normally incident light through arrays of
sub-wavelength holes, to a good approximation the resonance
wavelengths are given by a standing-wave quantization of the
surface plasmon dispersion for a smooth film at normal
incidence~\cite{Ebbesen:1998,Ghaemi:1998},
\begin{eqnarray}
\lambda^{SP}_{max}(i,j)=\frac{L}{\sqrt{i^2+j^2}}\sqrt{\frac{\varepsilon_1\varepsilon_2}{\varepsilon_1+\varepsilon_2}}.
\label{eqn}
\end{eqnarray}
Here, L is the periodicity of the array, $\varepsilon_1$ is the
dielectric constant of the interface medium, $\varepsilon_2$ is
that of the metal, and $i$ and $j$ are integers defining the
different diffraction orders. Certainly, each interface can
sustain SPs and the transmission spectra contain two sets of peaks
associated with each surface. In accordance with Eq.~(\ref{eqn}),
we conclude that the peak around 617 nm in Fig.~\ref{shapes}, due
to periodic surface plasmon wave, is related to the (1,0) Au-glass
resonance. For the Au-air resonant modes, the resonant wavelengths
are always less than 425 nm, which are not considered in this
paper. The peak related to the surface plasmon mode, hardly shifts
when varying the size of the air holes, which can be explained
well by Eq.~(\ref{eqn}). On the other hand, another resonant mode,
corresponding to the localized waveguide resonant mode
significantly depends on the hole size, which is naturally
understood by the resonant condition. Both results shown in
Fig.~\ref{shapes} coincide with those mentioned in
Ref.~\cite{Ruan:2006}, which gives another explanation for the
mechanism of the enhanced transmission.

\begin{figure}[t!]
\begin{center}
\epsfig{file=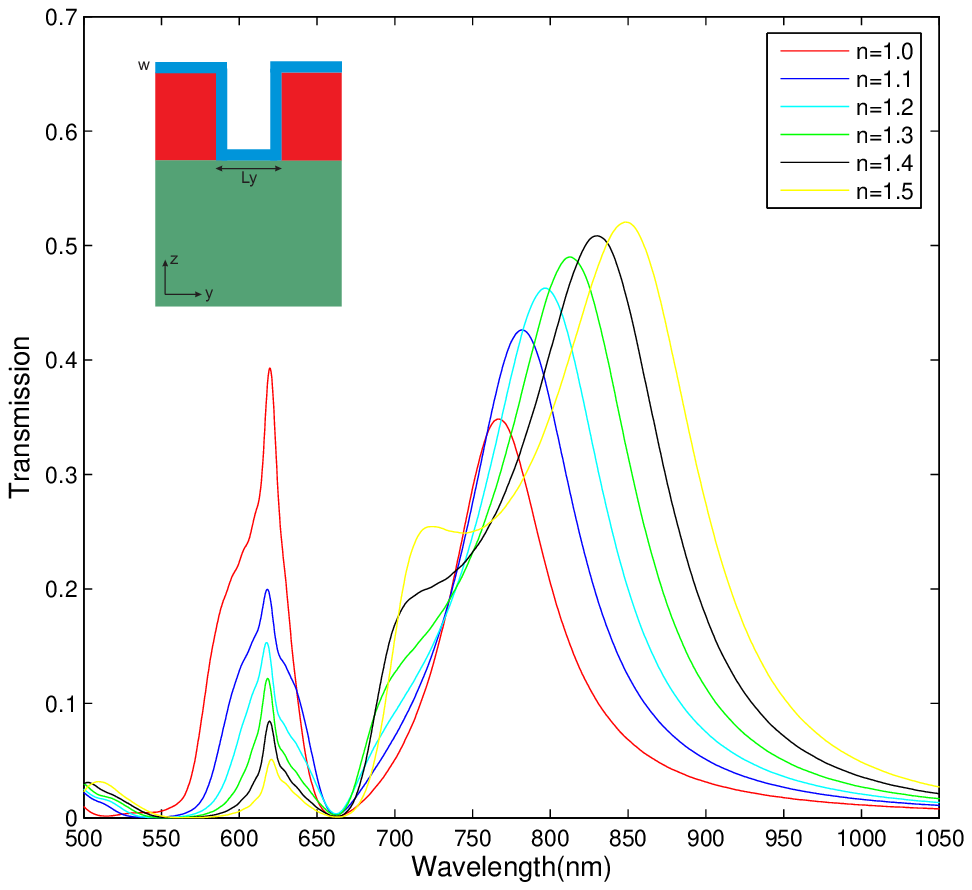,width=1 \columnwidth,clip,angle=0}
\end{center}
\caption{Transmission through the Au films with periodic arrays of
nanoholes being coated by different materials with refractive
indices varying from n=1.0 to 1.5 in steps of 0.1.}
\label{sensor1}
\end{figure}

\begin{figure}[t!]
\begin{center}
\epsfig{file=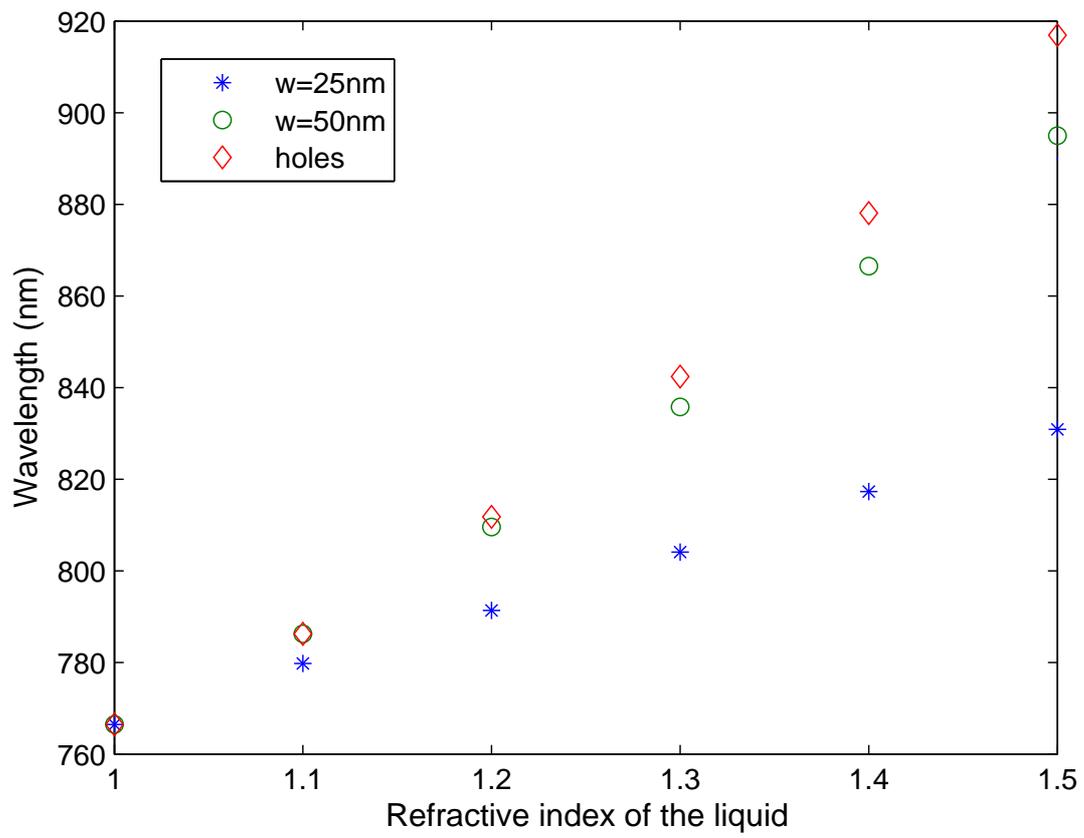,width=1 \columnwidth,clip,angle=0}
\end{center}
\caption{Wavelength shift as a function of the refractive index of
the material.} \label{sensitivity}
\end{figure}

As mentioned above, the response of the SP is very sensitive to
the refractive index in the vicinity of the metal surface. Hence,
simply placing a thin layer on a hole array will shift the
position of its transmission peaks. Consider a structure with yz
cross section shown in the inset of Fig.~\ref{sensor1}. The
dimensions of the holes in xy directions are $225\times150$ nm$^2$
and the lattice constant is 425 nm. Other parameters correspond to
those mentioned in Fig.~\ref{shapes}. Suppose the metal interface
is covered by a uniform layer (blue region), with a height of w=25
nm, see inset of Fig.~\ref{sensor1}. Figure~\ref{sensor1} shows
transmission spectra for the Au film being covered by a uniform
layer with a refractive index increasing from n=1.0 to n=1.5 in
steps of 0.1. One can see clearly that the peak, around 617 nm,
does not shift when varying the refractive index of the attached
layer. As seen from Eq.~(\ref{eqn}), the resonant peak (617 nm),
related to the (1,0) Au-glass resonance, only depends on the
lattice constant and the effective refractive index at the
metal-glass interface. Obviously, it is independent on the
coverage of the Au-air interface, i.e. this resonant mode is not a
candidate for a SPR sensor. We emphasize that the simulation
results agree well with theoretical ones obtained from
Eq.~(\ref{eqn}). It should be noted that the surface plasmon
resonant modes on the Au-air surface depend on the refractive
index of the layer attached to the surface, so that the SPR can be
used as a sensor. However, in this paper, we do not consider these
modes since they fall outside our wavelength region of interest.
On the other hand, peaks, corresponding to the localized waveguide
modes, significantly shift due to the change of the refractive
index of the layer. As mentioned above, each air hole can be
considered to be a section of metallic waveguide with both ends
open to free space, forming a low-quality-factor resonator. When
varying the effective refractive index of the layer in air holes,
the resonant condition changes and the corresponding peaks
obviously shift. The peak related to the localized waveguide mode
shifts $\delta\lambda$ =16 nm when the surface is modulated from
n=1.0 to n=1.1, thus giving potential for sensor applications.
When we further increase the height of the layer, the peak related
to the periodic surface plasmon wave hardly shifts while the peak
for the localized waveguide mode is strongly sensitive to the
refractive index of the layer. The sensitivity becomes larger when
increasing the height of the layer. As an example, the shift
becomes $\delta\lambda$ =18 nm for the case of w=50 nm as shown in
Fig. ~\ref{sensitivity}, when the surface is modified from n=1.0
to n=1.1. It was always believed that only periodic surface
plasmon resonant mode contributes to sensing for the SPR based
sensor. In this paper, we emphasize the contribution for sensing
from the localized waveguide mode, although the detected peak
position change is not significant and the peak is quite broad.
This may pave a new way to design SPR based sensors.

\begin{figure}[t!]
\begin{center}
\epsfig{file=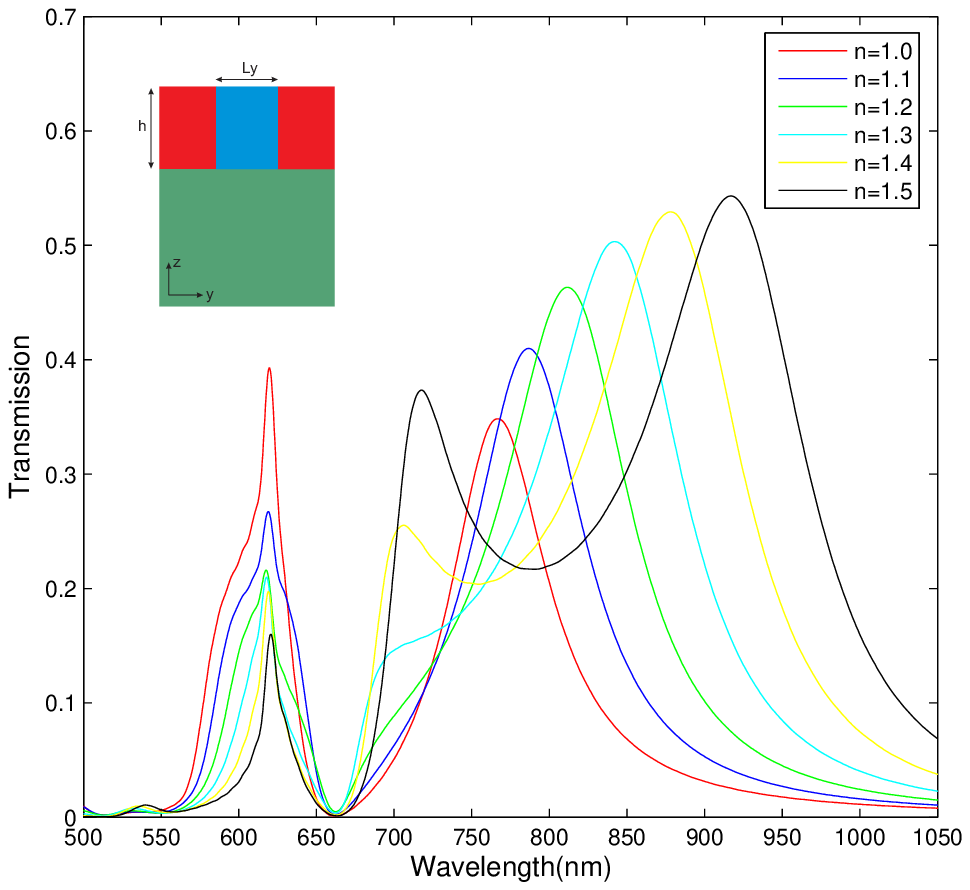,width=1 \columnwidth,clip,angle=0}
\end{center}
\caption{Transmission through the Au films with periodic arrays of
nanoholes being filled by different medias with refractive indices
varying from n=1.0 to 1.5 in steps of 0.1.} \label{sensor2}
\end{figure}

The analysis above demonstrates that the localized waveguide mode
takes an important role for the SPR based sensor. Furthermore, we
recently found that for frequency-independent dielectric-function
structure the sensitivity is proportional to the filling factor
$f$, defined by $f=\big<E \big|D\big>_l/\big<E
\big|D\big>$~\cite{Xiao:2006b}. The integral in the numerator of
the filling factor is restricted to the region containing the
fluid, or in the present case the dielectric layer, while the
other integral is over all space. The periodic surface plasmon
mode shown above is a (1,0) Au-glass resonance mode, where most of
the energy is believed to be bound at the Au-glass surface.
Therefore, $f$ is close to zero, i.e. the peak related to this
mode will not shift when the Au-air surface is modified, which is
in agreement with the calculated result in Fig.~\ref{sensor1}. To
increase the sensitivity, we further consider the structure with
air holes being filled with dielectric media (blue region). Other
parameters correspond to those in Fig.~\ref{sensor1}.
Figure~\ref{sensor2} shows transmission spectra for the Au film
with air holes being filled by different materials with the
refractive index increasing from n=1.0 to n=1.5 in steps of 0.1.
Similar to the result in Fig.~\ref{sensor1}, the peak around 617
nm in Fig.~\ref{sensor2} does not shift when varying the
refractive index of the media in the air holes, which agrees well
with that obtained from Eq.~(\ref{eqn}). On the other hand, the
peak related to the localized waveguide mode significantly shifts
due to the change of the refractive index of the media. As
mentioned above, each air hole can be considered to be a
low-quality-factor resonator. When varying the refractive index of
the media in the air holes, obviously the peak does shift.
Compared to the result shown in Fig.~\ref{sensor1}, the
sensitivity becomes better as shown in Fig. ~\ref{sensitivity}.
The shift for the localized waveguide mode is around
$\delta\lambda$ =20 nm when the air holes (with index n=1) are
filled by a madia of index n=1.1. Due to the low quality-factor,
$f$ will never be close to unity. If the quality factor of the
waveguide resonator can be increased, we believe that the
sensitivity will become stronger since $f$ will become larger. We
note that another peak starts to appear when the refractive index
increases above 1.3, which is also dependant on the refractive
index of the medias as shown in Fig.~\ref{sensor2}. Quite
naturally, the high-order mode appears, corresponding to a
high-order resonant waveguide mode appearing in a cavity composed
by metallic waveguide when increasing of the refractive index of
the material inside the cavity.

\section{Concluding remarks}
In this paper, we have studied transmission through arrays of
subwavelength holes in gold films utilizing the rigid
full-vectorial three-dimensional finite-difference time-domain
method. Based on the two different resonance mechanism for the
enhanced transmission, our calculations show that in our
wavelength range of interest, the localized waveguide resonant
mode attributes to sensing rather than periodic gold-glass surface
plasmon resonant modes. It was believed that only periodic surface
plasmon resonant mode contributes for sensing for the SPR based
sensor. In this paper, we emphasize the contribution for sensing
from the localized waveguide mode. Although the detected peak is
fairly broad and the shift is not too pronounced, this may pave a
new way to design of SPR based sensors.

\section*{Acknowledgments}
This work is financially supported by the Danish Council for
Strategic Research through the Strategic Program for Young
Researchers (grant no: 2117-05-0037). M. Qiu acknowledges the
support from the Swedish Foundation for Strategic Research (SSF)
on INGVAR program, the SSF Strategic Research Center in Photonics,
and the Swedish Research Council (VR).

\newpage

\end{document}